\documentclass[cameraready]{Interspeech}

\title{Learning Emotion-discriminative Representations for Zero-Shot Cross-lingual Speech Emotion Recognition}

\author[affiliation={1}, correspondingauthor]{Jinyi}{Mi}
\author[affiliation={1}]{Ding}{Ma}
\author[affiliation={2}]{Tomoki}{Toda}

\address{
    \vspace{-0.6em}
    $^1$ Graduate School of Informatics, Nagoya University, Japan \\
    $^2$ Information Technology Center, Nagoya University, Japan
}
\email{\{mi.jinyi,ding.ma\}@g.sp.m.is.nagoya-u.ac.jp,tomoki@icts.nagoya-u.ac.jp}

\usepackage{comment}
\usepackage{algorithm}
\usepackage{algorithmic}
\usepackage{multirow}
\usepackage{makecell}
\usepackage{amsmath}

\begin{document}

\maketitle

\begin{abstract}
    Zero-shot cross-lingual speech emotion recognition (SER) remains challenging due to distribution mismatches across languages and the lack of emotion annotations in target language. Under such conditions, models trained solely on source-language data frequently suffer from degraded generalization when evaluated on unseen target languages. To address this limitation, we propose an emotion-discriminative representation learning method that integrates supervised contrastive learning and speaker adversarial learning. The contrastive learning promotes cross-lingual emotion alignment, while speaker adversarial learning suppresses speaker-related cues to encourage speaker-invariant representations. Experimental results under a zero-shot cross-lingual SER setting demonstrate that the proposed method significantly improves SER performance over conventional training strategies.
\end{abstract}

\section{Introduction}

Speech Emotion Recognition (SER) has received increasing attention in affective computing over the past two decades~\cite{Schuller18_Review} owing to its potential applications in healthcare~\cite{Uddin20_healthcare, Hossain15_healthcare}, education~\cite{Li07_education, Tickle13_education}, and robotics~\cite{Gamboa24_robot, 25_robot}. SER aims to identify human emotional states by extracting emotional features from speech. Research on SER systems has shown good performance when the training and test data are from the same language~\cite{Mi2024two, Chen2023Exploring, MI2026101987}. However, when evaluated on speech data from a different language, their performance often degrades significantly~\cite{latif2018_cross, Neumann2018CRoss}, limiting real-world practicality. Therefore, cross-lingual SER has emerged as a crucial topic. It refers to the setting where an SER model trained on a \textit{source} language is expected to generalize to a different \textit{target} language. Nevertheless, developing robust cross-lingual SER systems remains challenging due to linguistic and cultural variations in emotional expression across languages. 

Most approaches~\cite{Ahn2021cross, Han2025Cross} address cross-lingual SER through transfer learning to mitigate domain shift between languages. For example, Self-Supervised Learning (SSL) models like wav2vec 2.0~\cite{baevski2020wav2vec} and WavLM~\cite{chen2022wavlm} are typically pretrained on large-scale speech corpora from source-language data and then fine-tuned on target-language emotional data to adapt to language-specific characteristics. However, such transfer learning methods generally require sufficient labeled training data in the target language to achieve satisfactory cross-lingual performance. In practice, emotional speech datasets in many languages are low-resource and often lack reliable emotion annotations, making conventional transfer learning less feasible. 

In light of this, Latif~et~al.~\cite{latif2019unsupervised} propose an unsupervised Generative Adversarial Network (GAN)-based~\cite{goodfellow2014generative} framework to learn language-invariant representations for cross-lingual SER. Similarly, Cai~et~al.~\cite{cai2021unsupervised} introduce a Domain Adversarial Neural Network (DANN)~\cite{ganin2016domain}-based model with a language classifier and a Gradient Reversal Layer (GRL) to alleviate cross-lingual distribution mismatch. Additionally, Upadhyay~et~al.~\cite{upadhyay24_layer} propose a layer-anchoring strategy that leverages the hierarchical structure of speech models to find the commonality between corpora, facilitating cross-lingual emotion transfer. As another approach, Tang~et~al.~\cite{tang2023end} propose a transfer learning approach by employing multilingual SSL pretraining for downstream cross-lingual SER. Consequently, cross-lingual generalization in this approach is largely attributed to sufficiently large-scale multilingual pretraining over 50 languages. Furthermore, although the aforementioned methods do not require target-language emotion labels, they rely on some form of target-language information, such as access to speech data or language labels from the target language during training or multilingual pretraining. Additionally, these studies mainly focus on reducing cross-lingual distribution mismatch, rather than explicitly modeling emotion-level structural consistency across languages. As a result, the alignment of emotion categories across languages may not be sufficiently enforced, which can limit cross-lingual generalization.

\vspace{-0.2mm}
Motivated by these observations, we propose an emotion-discriminative representation learning method to address the problem of zero-shot cross-lingual SER. Specifically, our method integrates supervised contrastive learning and speaker adversarial learning to learn emotion-relevant, language-invariant, and speaker-invariant representations, enabling the model to capture emotion-level cross-lingual structural consistency from a few language datasets. We conduct systematic experiments under zero-shot cross-lingual settings to evaluate our method. In addition, we visualize the learned representation space to examine the clustering behavior of emotion categories. Our contributions to this work are as follows:
\vspace{-0.9mm}
\begin{itemize}
\item We propose a framework for zero-shot cross-lingual SER that explicitly aligns emotion-conditioned representations across languages through supervised contrastive learning.
\vspace{-0.5mm}
\item We incorporate speaker adversarial learning to disentangle speaker variability, facilitating the learning of speaker-invariant emotion representations.
\vspace{-0.5mm}
\item We evaluate the proposed method under nine zero-shot cross-lingual settings and visualize the learned representations. The results demonstrate improved cross-lingual generalization performance cost-efficiently via the training data with only a few languages. Furthermore, ablation studies demonstrate the benefits of supervised contrastive learning and speaker adversarial learning to the proposed method.
\end{itemize}
\vspace{-0.5mm}
\section{Methodology}

\begin{figure}[t]
\centering
\includegraphics[width=1.0\linewidth]{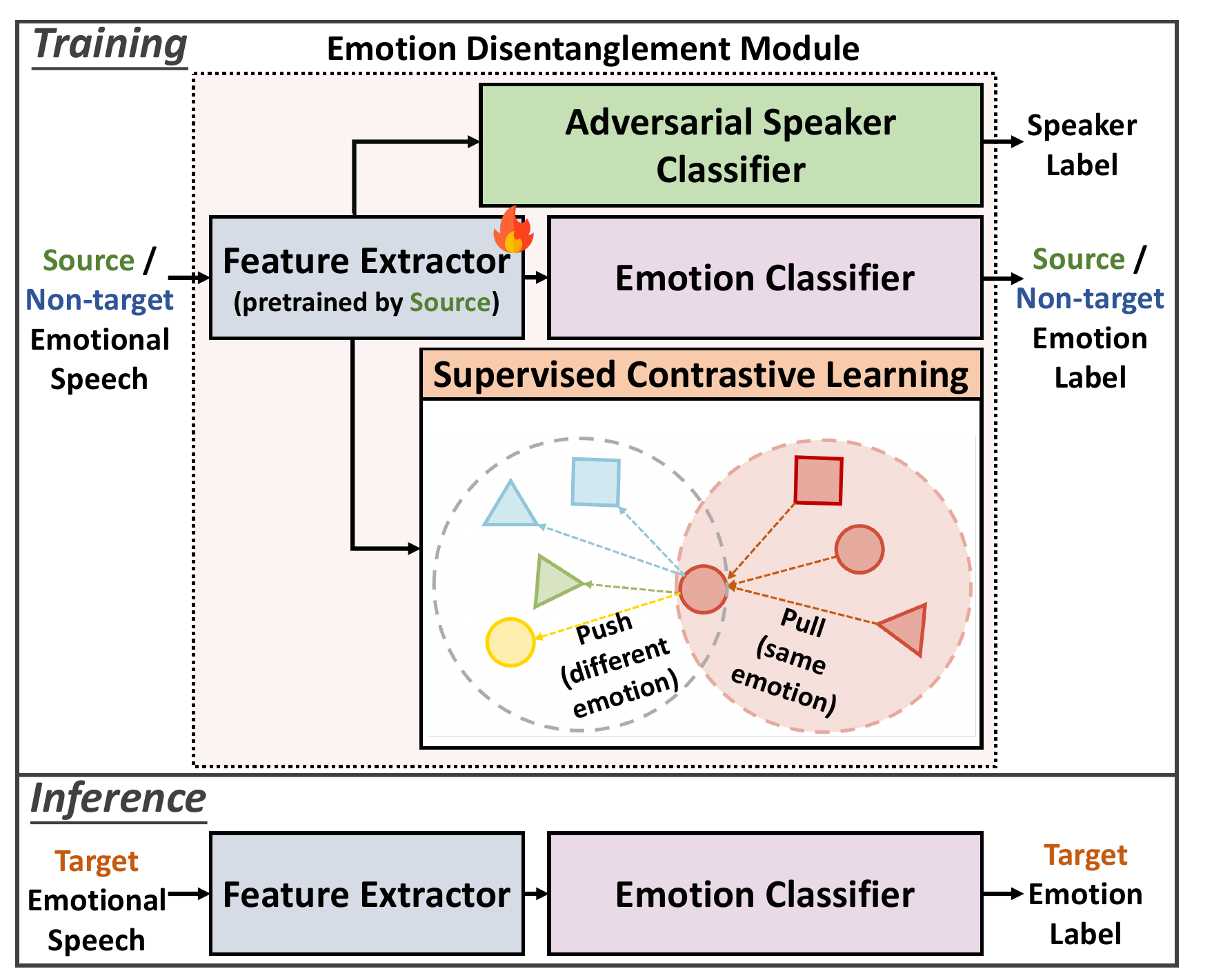}
\setlength{\abovecaptionskip}{-6pt}
\caption{Illustration of the proposed method. “Source” and “Target” denote the source and target languages, respectively. “Non-target” denotes non-target languages.}
\label{fig:Overview_framework}
\vspace{-7.5mm}
\end{figure}

Let \(\mathcal{D}_\text{source}\) denote the labeled emotional speech data of the source language, \(\mathcal{D}_\text{non-target}\) the union of labeled emotional speech data from auxiliary non-target languages, and \(\mathcal{D}_\text{target}\) the emotional speech data in the target language. Note that \(\mathcal{D}_\text{target}\) is not used during training. The overall training data is defined as
\begin{align}
\mathcal{D} = \mathcal{D}_\text{source} \cup \mathcal{D}_{\text{non-target}}.
\end{align}
Each training sample is represented as
\begin{align}
(\boldsymbol{x},y, \ell, s) \in \mathcal{D},
\end{align}
where \(\boldsymbol{x}\) denotes an emotional speech utterance, \(y \in \mathcal{Y}\) represents its emotion label,  \(\ell\ \in \mathcal{G}\) denotes the language label, and \(s\) is the speaker identity. Here, \(\mathcal{G}\) denotes the set of languages in the training data, and \(\mathcal{Y}\) is the shared emotion label space with \(C\) predefined emotion classes.

As illustrated in Figure~\ref{fig:Overview_framework}, the overall framework can be viewed as an emotion representation disentanglement paradigm with feature extractor, supervised contrastive learning, adversarial speaker learning, and emotion classification. Details of these modules are provided in the following subsections.

\vspace{-3mm}
\subsection{Feature Extraction Module}
\label{sec:feature_extractor}
\vspace{-1mm}
We utilize a pretrained SSL model, wav2vec 2.0, as the feature extractor, which is pretrained on a common speech corpus from the source language. wav2vec 2.0 integrates Convolution Neural Network (CNN) layers with a Transformer encoder to effectively obtain contextualized speech representations:
\begin{align}
\boldsymbol{H} = \mathrm{wav2vec2}(\boldsymbol{x}) \in \mathbb{R}^{n \times d},
\end{align}
where \(n\) is the number of time frames, and \(d\) is the dimension of hidden representations. Then, we compute the mean-pooled vectors for the speech representations:
\begin{align}
\boldsymbol{h} = \mathrm{MeanPooling}(\boldsymbol{H}) \in \mathbb{R}^{d}.
\end{align}
\vspace{-8mm}
\subsection{Supervised Contrastive Learning}
\label{sec:supclr}
\vspace{-1mm}

We introduce a supervised contrastive learning method. This method aims to bring features belonging to the same emotion class closer and push features from different classes farther apart, which can promote the focus on emotion-related features. Moreover, we further introduce a language-aware weighting strategy into the supervised contrastive objective to explicitly strengthen cross-lingual emotion alignment. Unlike standard supervised contrastive learning treats all same-emotion samples equally, we assign larger weights to same-emotion pairs that share the same emotion label but belong to different languages.

Let \(I\) denote the index set of samples in each batch of training data. For each anchor \(i \in I\), the same emotion set is defined as
\begin{align}
P(i) = \{ p \in I \mid y_p = y_i,\, p \neq i \},
\end{align}
and the set of all non-anchor samples is defined as
\begin{align}
A(i) = I \setminus\{i\}.
\end{align}
We denote the language label of the anchor sample \(i\) as \(\ell_i\). The language-aware weight assigned to a same-emotion pair \((i,p)\) is defined as
\vspace{-1mm}
\begin{align}
w_{i,p} =
\begin{cases}
\lambda, & \ell_p \neq \ell_i, \\
1, & \ell_p = \ell_i,
\end{cases}
\end{align}
where \(\lambda>1\) is a hyperparameter controlling the strength of cross-lingual alignment. The supervised contrastive learning loss \(\mathcal{L}_{\mathrm{SupCLR}}\) is formulated as

\begin{align}
\mathcal{L}_{\mathrm{SupCLR}}
&=
-\frac{1}{|I|}
\sum_{i \in I}
\frac{
\sum_{p \in P(i)}
w_{i,p}
\log
\frac{
\exp\left(\phi(\boldsymbol{h}_i, \boldsymbol{h}_p)/\tau\right)
}{
\sum_{a \in A(i)}
\exp\left(\phi(\boldsymbol{h}_i, \boldsymbol{h}_a)/\tau\right)
}
}{
\sum_{p \in P(i)} w_{i,p}
}.
\end{align}
where \(\phi(\cdot,\cdot)\) denotes cosine similarity and \(\tau\) is a temperature parameter. \(i\), \(p\), \(a\) denote the indices of the anchor, same-emotion, and non-anchor samples, respectively.

Due to imbalance of data across different languages, we design a hierarchical cross-lingual sampling strategy to construct the index set $I$ for supervised contrastive learning. We first sample $N_{\text{lang}}$ distinct languages from $\mathcal{G}$, and then sample $N_{\text{cls}}$ emotion classes from the shared emotion label space $\mathcal{Y}$, where $N_{\text{lang}} \ge 2$ and $N_{\text{cls}} \ge 2$. For each selected language--emotion pair, $N_{\text{sam}}$ instances are randomly drawn and added to $I$, where $N_{\text{sam}} \ge 2$. Consequently, the constructed index set $I$ contains $N_{\text{lang}} \times N_{\text{cls}} \times N_{\text{sam}}$ samples, ensuring that multiple languages and emotion classes coexist within $I$. This facilitates the formation of both intra-class and cross-lingual contrastive pairs.

\vspace{-3mm}
\subsection{Adversarial Speaker Learning}
\label{sec:spkadv_classifier}
To prevent the model from exploiting speaker-related characteristics as a shortcut for emotion classification, we further introduce a speaker adversarial classifier to disentangle speaker identity in emotional representations. The classifier consists of a GRL followed by two linear layers with a ReLU activation and a dropout layer in between. This process is formulated as
\begin{align}
\hat{\boldsymbol{s}}
=
\mathrm{Linear}(
\mathrm{Dropout}(
\mathrm{ReLU}(
\mathrm{Linear}(
\mathrm{GRL}(\boldsymbol{h})
)
)
)
).
\end{align}
Let $s_i$ denote the ground-truth speaker identity of sample $i\in I$. The speaker adversarial loss is formulated as
\begin{align}
\mathcal{L}_{\mathrm{SpkAdv}}
=
-\frac{1}{|I|}
\sum_{i \in I}
\log (\mathrm{Softmax}(\hat{\boldsymbol{s}}_i))_{s_i},
\end{align}
where $\hat{\boldsymbol{s}}_i$ denotes the predicted speaker logits, and $(\mathrm{Softmax}(\hat{\boldsymbol{s}}_i))_{s_i}$ is the softmax probability of the correct speaker label. Due to GRL, the speaker classifier is trained to minimize \(\mathcal{L}_{\mathrm{SpkAdv}}\), while the feature extractor is trained adversarially to maximize it.

\vspace{-2mm}
\subsection{Emotion Classifier}
\label{sec:emo_classifier}
Emotion classification is performed with a linear layer followed by a \(\mathrm{Softmax}\) activation to the speech embedding \(\boldsymbol{h}\), i.e.,
\begin{align}
\pi(y|\boldsymbol{h})
=
\mathrm{Softmax}(\mathbf{W}_{c}\boldsymbol{h}+\boldsymbol{b}_{c})
\in \mathbb{R}^{C},
\end{align}
where $C$ is the number of emotion classes, and $\mathbf{W}_{c}$ and $\boldsymbol{b}_{c}$ are learnable parameters.

\vspace{-2.5mm}
\subsection{Training Objective}
\label{sec:loss_function}
\vspace{-1mm}
During training, the parameters of the feature extractor, emotion classifier, and speaker classifier are jointly optimized. The cross-entropy (CE) loss is adopted for emotion classification, while the supervised contrastive loss and the adversarial speaker loss are defined in Sections~\ref{sec:supclr} and ~\ref{sec:spkadv_classifier}, respectively. The overall loss is formulated as
\begin{align}
\mathcal{L} = \mathcal{L}_{\mathrm{CE}} + \alpha \mathcal{L}_{\mathrm{SupCLR}} + \beta \mathcal{L}_{\mathrm{SpkAdv}},
\end{align}
where \(\alpha\) and \(\beta\) are the hyperparameters for balancing the weight of the three loss terms.

\vspace{-2mm}
\section{Experimental Evaluations}
\subsection{Datasets and Evaluation Metrics}
For clarity, we use \textit{EN}, \textit{CN}, \textit{DE}, \textit{FR}, \textit{UR} to denote English, Mandarin, German, French, and Urdu, respectively. We design nine zero-shot cross-lingual settings. In each setting, one language is selected as the source and another as the target, while the rest serve as non-target languages. The configuration of each setting is summarized in Table~\ref{tab:settings}. We describe the employed datasets below. Note that we focus on four emotion classes: \textit{happy}, \textit{angry}, \textit{sad}, and \textit{neutral} for all datasets. 

{\par\noindent\textbf{MELD:} The Multimodal EmotionLines Dataset (MELD)~\cite{poria2019meld} contains over 13,000 English utterances from the TV series \textit{Friends}, involving 304 speakers. It is annotated with seven emotions (angry, disgust, sad, happy, neutral, surprise, and fear). We followed the official train/validation/test split.}

{\par\noindent\textbf{ESD:} The Emotional Speech Database (ESD) corpus~\cite{Zhou2022ESD} contains about 29-hour recordings from 10 English and 10 Mandarin speakers. We used the Mandarin subset with train/validation/test set of 11,200/1,400/1,400 utterances.}

{\par\noindent\textbf{EMO-DB:} It is a German emotional corpus containing over 700 utterances from 10 actors, labeled with seven emotions (angry, boredom, disgust, fear, happy, neutral, and sad)~\cite{burkhardt05b_emodb}. We split it into train/validation/test sets with 266/35/38 utterances.}

{\par\noindent\textbf{CaFE:} The Canadian French Emotional (CaFE)~\cite{gournay2018cafe} contains 936 French emotional utterances spoken by 12 speakers, labeled with seven emotions (angry, disgust, happy, sad, surprise, neutral, and fear). We split it into train/validation/test sets with 420/42/42 utterances.}

{\par\noindent\textbf{URDU:} It involves Urdu emotional recordings from the Urdu TV talk shows~\cite{latif2018urdu} with 400 utterances from 38 speakers, annotated with four emotions (angry, happy, sad, and neutral). We split it into train/validation/test sets with 300/20/80 utterances.}

The evaluation metrics are Unweighted Average Recall (UAR) and Macro-F1 (F1), which are widely used for cross-lingual and class-imbalanced classification tasks~\cite{latif2019unsupervised, cai2021unsupervised, MI2026101987, shi2024study, Mi2024two}.
\begin{table}[t]\footnotesize
\caption{Configuration of cross-lingual training settings. “\#Samples” indicates the number of training samples, “\#Spk” is the number of speakers.}
\label{tab:settings}
\vspace{-1.5mm}
\centering
\renewcommand{\arraystretch}{1.0}
\setlength{\tabcolsep}{3pt}
\begin{tabular}{cccccc}
\toprule[1pt]
\textbf{\makecell{Cross-lingual\\Task}} & \textbf{Source} & \textbf{Non-target} & \textbf{Target} & \textbf{\#Samples} & \textbf{\#Spk} \\ \midrule
\textbf{EN$\rightarrow$DE} & EN& CN, UR, FR& DE& 20165& 288\\
\textbf{CN$\rightarrow$DE} & CN& EN, UR, FR& DE& 20165& 288\\
\textbf{FR$\rightarrow$DE} & FR& EN, CN, UR& DE& 20165& 288\\
\textbf{EN$\rightarrow$FR} & EN& CN, DE, UR& FR& 20011& 286\\
\textbf{CN$\rightarrow$FR} & CN& EN, DE, UR& FR& 20011& 286\\
\textbf{DE$\rightarrow$FR} & DE& EN, CN, UR& FR& 20011& 286\\
\textbf{EN$\rightarrow$CN} & EN& DE, UR, FR& CN& 9231& 288\\
\textbf{DE$\rightarrow$CN} & DE& EN, UR, FR& CN& 9231& 288\\
\textbf{FR$\rightarrow$CN} & FR& EN, DE, UR& CN& 9231& 288\\
\bottomrule[1pt]
\end{tabular}
\vspace{-6mm}
\end{table}

\begin{table*}[!t]\footnotesize
\caption{Comparison of the proposed, baseline, and upper bound systems using UAR (\%) and F1 (\%).}
\label{tab:main_results}
\vspace{-1.5mm}
\centering
\renewcommand{\arraystretch}{1.1}
\setlength{\tabcolsep}{4pt}
\begin{tabular}{ccccccccccc|cc}
\toprule[1pt]
\multirow{2}{*}{\textbf{\makecell{Cross-lingual\\Task}}} &
  \multicolumn{2}{c}{Baseline 1} &
  \multicolumn{2}{c}{Baseline 2} &
  \multicolumn{2}{c}{Proposed} &
  \multicolumn{2}{c}{\makecell{Proposed\\(w/o $\mathcal{L}_{\mathrm{SpkAdv}}$)}} &
  \multicolumn{2}{c}{\makecell{Proposed\\(w/o $\mathcal{L}_{\mathrm{SupCLR}}$)}} &
  \multicolumn{2}{!{\hspace{-0.34pt}\vrule height 3.4ex depth 1.2ex width 0.4pt}c}{Upper Bound} \\
\addlinespace[1.5pt] \cline{2-13} \addlinespace[1.8pt]
& \makecell{UAR ($\uparrow$)} & \makecell{F1($\uparrow$)}
& \makecell{UAR ($\uparrow$)} & \makecell{F1 ($\uparrow$)}
& \makecell{UAR ($\uparrow$)} & \makecell{F1 ($\uparrow$)}
& \makecell{UAR ($\uparrow$)} & \makecell{F1 ($\uparrow$)}
& \makecell{UAR ($\uparrow$)} & \makecell{F1 ($\uparrow$)}
& \makecell{UAR ($\uparrow$)} & \makecell{F1 ($\uparrow$)} \\
\midrule
\textbf{EN$\rightarrow$DE}                    & 52.23 & 43.11       & 88.19 & 88.68       & \textbf{94.64} & \textbf{94.36}       & 92.66 & 93.00       & 90.87 & 91.81       & 97.22 & 97.67       \\
\textbf{CN$\rightarrow$DE}                    & 85.42 & 86.66       & 88.54 & 89.25       & \textbf{94.44} & \textbf{95.21}       & 91.67 & 92.58       & 88.89 & 89.73       & 97.22 & 97.06       \\
\textbf{FR$\rightarrow$DE}                    & 73.91 & 75.17       & 80.95 & 80.29       & \textbf{88.89} & \textbf{89.73}       & 86.31 & 86.31       & 83.53 & 83.67       & 95.44 & 95.44       \\

\textbf{EN$\rightarrow$FR}                    & 45.83 & 39.81       & 70.83 & 68.04       & \textbf{77.08} & \textbf{75.35}       & 75.00 & 74.41       & 72.92 & 70.59       & 87.50 & 84.42       \\
\textbf{CN$\rightarrow$FR}                    & 72.92 & 75.74       & 79.17 & 77.21       & \textbf{85.42} & \textbf{86.86}       & 83.33 & 85.08       & 79.17 & 80.43       & 89.58 & 91.14       \\
\textbf{DE$\rightarrow$FR}                    & 47.92 & 47.96       & 68.75 & 68.81       & \textbf{75.00} & \textbf{72.84}       & 72.92 & 71.64       & 70.83 & 70.49       & 81.25 & 78.00       \\

\textbf{EN$\rightarrow$CN}                    & 58.36 & 56.19       & 71.14 & 70.90       & \textbf{78.07} & \textbf{77.92}       & 76.21 & 75.17       & 73.21 & 72.48       & 97.07 & 97.08       \\
\textbf{DE$\rightarrow$CN}                    & 48.79 & 50.11       & 53.64 & 53.07       & \textbf{73.86} & \textbf{73.48}       & 71.71 & 71.98       & 64.93 & 64.18       & 91.79 & 91.70      \\
\textbf{FR$\rightarrow$CN}                    & 50.07 & 49.38       & 57.64 & 56.98       & \textbf{72.93} & \textbf{71.87}       & 71.21 & 71.09       & 67.36 & 66.89       & 90.21 & 90.21       \\ \midrule
\textbf{Avg.}        & 59.49 & 58.24       & 73.21 & 72.58       & \textbf{82.26} & \textbf{81.96}       & 80.11 & 80.14       & 76.86 & 76.70       & 91.92 & 91.41       \\
\bottomrule[1pt]
\end{tabular}
\vspace{-4mm}
\end{table*}

\vspace{-6mm}
\subsection{System Building}
\vspace{-1mm}
We prepared two baselines, namely \textit{Baseline 1} and \textit{Baseline 2}, and three variants of our proposed method, referred to as \textit{Proposed}, \textit{Proposed w/o \(\mathcal{L}_{\mathrm{SupCLR}}\)}, and \textit{Proposed w/o \(\mathcal{L}_{\mathrm{SpkAdv}}\)}, respectively. Furthermore, we included an \textit{Upper bound} system that used target-language supervision. All systems share the same feature extractor and emotion classifier architecture, with differences arising from training data and the incorporation of supervised contrastive and speaker adversarial learning.

\begin{itemize}
\item \textit{Baseline 1}: A standard cross-lingual SER baseline as in~\cite{Han2025Cross}, consisting of a pretrained feature extractor and emotion classifier architecture, fine-tuned using the source-language data. 

\item \textit{Baseline 2}: The same architecture as \textit{Baseline 1}, fine-tuned using the data of source language and non-target languages.

\item \textit{Upper bound}: The same architecture as the baselines, fine-tuned using the target-language data for the performance ceil. Note that \textit{Upper bound} is not a zero-shot setting.

\item \textit{Proposed}: Our full model with $\mathcal{L}_{\mathrm{SupCLR}}$ and $\mathcal{L}_{\mathrm{SpkAdv}}$.

\item \textit{Proposed w/o \(\mathcal{L}_{\mathrm{SupCLR}}\)}: An ablation variant that removes the supervised contrastive objective.

\item \textit{Proposed w/o \(\mathcal{L}_{\mathrm{SpkAdv}}\)}: An ablation variant that removes the speaker-adversarial objective.
\end{itemize}

\vspace{-3mm}
\subsection{Implementation}
\vspace{-1mm}
Our method was implemented using Python 3.10 and Pytorch 2.0.1. Training and evaluation were conducted on a system equipped with an Intel(R) Xeon(R) Gold 6248 CPU (2.50 GHz), 32 GB RAM, and one NVIDIA Tesla V100 GPU. For the feature extractor, the speech encoder was initialized using language-matched pretrained wav2vec~2.0 Base models for each source language
(\textit{EN}\footnote{\href{https://huggingface.co/facebook/wav2vec2-base-960h}{\nolinkurl{facebook/wav2vec2-base-960h}}}, \textit{CN}\footnote{\href{https://huggingface.co/TencentGameMate/chinese-wav2vec2-base}{\nolinkurl{TencentGameMate/chinese-wav2vec2-base}}}, \textit{DE}\footnote{\href{https://huggingface.co/facebook/wav2vec2-base-de-voxpopuli-v2}{\nolinkurl{facebook/wav2vec2-base-de-voxpopuli-v2}}}, \textit{FR}\footnote{\href{https://huggingface.co/facebook/wav2vec2-base-fr-voxpopuli}{\nolinkurl{facebook/wav2vec2-base-fr-voxpopuli}}}). The wav2vec~2.0 encoder was fine-tuned using low-rank adaptation~\cite{hu2022lora}, bottleneck adaptor~\cite{houlsby2019parameter}, and weight gating~\cite{lashkarashvili2024parameter}. To ensure a fair and consistent comparison, the implementation was identical across all systems. During training, we set $\lambda=2.5$, $\alpha=1.0$, and $\beta=0.3$. For hierarchical batch sampling, we used $N_{\text{lang}}=3$, $N_{\text{cls}}=4$, and $N_{\text{sam}}=3$.

\begin{figure*}[!t]
\centering
\includegraphics[width=0.85\linewidth]{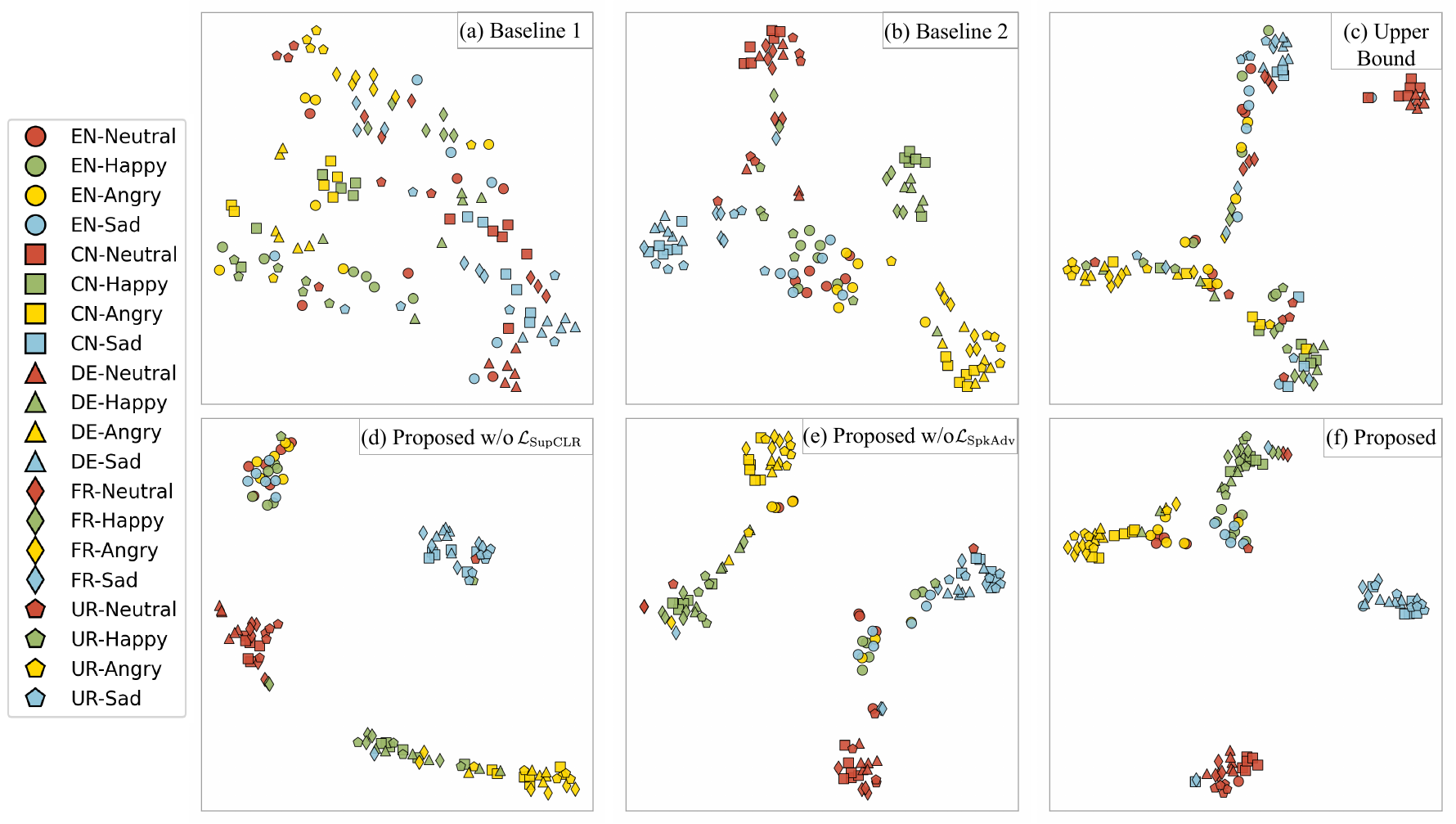}
\setlength{\abovecaptionskip}{-2pt}
\caption{Visualizations of hidden representations extracted from (a) Baseline 1, (b) Baseline 2, (c) Upper Bound, (d) Proposed w/o $\mathcal{L}_{\mathrm{SupCLR}}$, (e) Proposed w/o $\mathcal{L}_{\mathrm{SpkAdv}}$, and (f) Proposed, where colors represent emotion classes and marker shapes indicate languages.}
\label{fig:visualization}
\vspace{-4.5mm}
\end{figure*}

\vspace{-2.5mm}
\subsection{Experimental Results}
\vspace{-1mm}
{\par\noindent\textbf{Comparisons with baselines.} We compared the performance of proposed methods and baselines in nine cross-lingual settings. As shown in Table~\ref{tab:main_results}, all the proposed systems consistently outperform all baseline systems. \textit{Proposed} achieves the best performance compared to all other systems, where the average UAR and F1 reach 82.26\% and 81.96\%, respectively, achieving the closest performance to \textit{Upper bound}. Compared with \textit{Baseline 2}, \textit{Proposed} improves average UAR and F1 by 9.05\% and 9.38\%, respectively. Moreover, even using only a single component, i.e., either \textit{Proposed w/o $\mathcal{L}_{\mathrm{SpkAdv}}$} or \textit{Proposed w/o $\mathcal{L}_{\mathrm{SupCLR}}$}, still achieves better overall performance than the baselines. These findings demonstrate the effectiveness of our method for cross-lingual SER.

{\par\noindent\textbf{Ablation study.} We further evaluate the contribution of supervised contrastive and speaker-adversarial learning in our framework. Results in Table~\ref{tab:main_results} show that removing either component causes a performance drop, emphasizing that both objectives contribute to improved cross-lingual SER. Among the two, supervised contrastive learning has the largest impact, with its removal resulting in a 5.40\% UAR and 5.26\% F1 decrease, suggesting that its importance for cross-lingual generalization. Speaker adversarial learning is also beneficial, with a 2.15\% UAR and 1.82\% F1 drop when removed, highlighting its contribution to robustness against speaker variability.

{\par\noindent\textbf{Visualization analysis.} We employed t-SNE~\cite{van2008visualizing} to visualize the learned representations, as shown in Figure~\ref{fig:visualization}. We applied $\mathit{EN}\mkern-8mu\rightarrow\mkern-8mu\mathit{DE}$ task and randomly selected 6 utterances per emotion class from each language in the test set. Compared with \textit{Baseline 1}, \textit{Baseline 2} shows improved clustering, but the separation between emotion classes remains limited. \textit{Upper bound} presents a clear clustering effect for target language, but the representation space for other languages exhibits poor discriminability. In contrast, the three proposed systems not only achieve the clear clustering effects for the target language that are comparable to \textit{Upper bound}, but also yield more compact clusters for the other languages than both the baselines and \textit{Upper bound}. Furthermore, \textit{Proposed} achieves the most compact and well-separated clusters among all systems. These findings suggest that our method learns more emotion-discriminative representations and improves cross-lingual emotion alignment.}

\vspace{-3mm}
\section{Conclusion}
\vspace{-2mm}
In this paper, we propose an emotion-discriminative representation learning method that effectively integrates supervised contrastive learning and speaker adversarial learning. Extensive experiments under nine zero-shot cross-lingual settings demonstrate that the proposed systems significantly outperform baseline systems. Visualization evidence also reflects that our method can recognize generalized emotional representations across languages. Future work will explore (1) stronger cross-lingual alignment methods, and (2) incorporating multi-modal information to further approach the upper-bound performance.

\section{Acknowledgments}
This work was partly supported by JST CREST Grant Number JPMJCR22D1 and JSPS KAKENHI Grant Number 26H02530, Japan. In addition, this work was also financially supported by JST SPRING, Grant Number JPMJSP2125. The author would like to take this opportunity to thank the “THERS Make New Standards Program for the Next Generation Researchers.”

\section{Generative AI Use Disclosure}
Generative AI tools were used for grammar correction.

\bibliographystyle{IEEEtran}
\bibliography{main}

\end{document}